\documentclass[%
 reprint,
 superscriptaddress,
  amsmath,amssymb, 
  aps,
prl,
]{revtex4-2}

\usepackage{graphicx}
\usepackage{dcolumn}
\usepackage{bm}
\usepackage{hyperref}
\usepackage[mathlines]{lineno}
\usepackage{booktabs} 
\usepackage{color}
\usepackage{multirow}
\usepackage{upgreek}

\begin{document}

\preprint{APS/123-QED}

\title{Acceleration of Positive Muons by a Radio Frequency Cavity}
\newcommand{\JAEA}{Japan Atomic Energy Agency (JAEA), Tokai, Naka, Ibaraki 319-1195, Japan}
\newcommand{\KEK}{High Energy Accelerator Research Organization, Ibaraki 319-1106, Japan}
\newcommand{\UBC}{Laboratory for Advanced Spectroscopy and Imaging Research (LASIR), Department of Chemistry, University of British Columbia, Vancouver, BC, V6T 1Z1, Canada}
\newcommand{\IbarakiUniv}{Graduate School of Science and Engineering, Ibaraki University, Mito, Ibaraki 310-8512, Japan}
\newcommand{\OkayamaUniv}{Research Institute for Interdisciplinary Science, Okayama University, Okayama 700-8530, Japan}
\newcommand{\KyushuUniv}{Faculty of Science, Kyushu University, Fukuoka, Fukuoka 819-0395, Japan}
\newcommand{\RCAPP}{Research Center of Advanced Particle Physics, Kyushu University, Fukuoka, Fukuoka 819-0395, Japan}
\newcommand{\NiigataUniv}{Institute of Science and Technology, Niigata University, Niigata 950-2181, Japan}
\newcommand{\NagoyaUniv}{Graduate School of Science, Nagoya University, Nagoya, Aichi 464-8602, Japan}
\newcommand{\KMI}{Kobayashi-Maskawa Institute for the Origin of Particles and the Universe, Nagoya University, Nagoya, Aichi 464-8602, Japan}
\newcommand{\UnivofTokyo}{Graduate School of Science, University of Tokyo, 7-3-1 Hongo, Bunkyo-ku, Tokyo 113-0033, Japan}
\newcommand{\PekingUniv}{School of Physics, Peking University, Beijing 100871, China}
\newcommand{\SKLNPT}{State Key Laboratory of Nuclear Physics and Technology, Peking University, Beijing 100871, China}
\newcommand{\RIKEN}{Nishina Center for Accelerator-based Science, RIKEN, Wako, Saitama 351-0198, Japan}
\newcommand{\victoria}{Department of Physics and Astronomy, University of Victoria, Victoria, BC V8P 5C2, Canada}
\newcommand{\TRIUMF}{TRIUMF, Vancouver, BC V6T 2A3, Canada}

\author{S.~Aritome} \affiliation{\UnivofTokyo}
\author{K.~Futatsukawa} \affiliation{\KEK}
\author{H.~Hara} \affiliation{\OkayamaUniv}
\author{K.~Hayasaka} \affiliation{\NiigataUniv}
\author{Y.~Ibaraki} \affiliation{\NagoyaUniv}
\author{T.~Ichikawa} \affiliation{\NagoyaUniv}
\author{T.~Iijima} \affiliation{\NagoyaUniv}\affiliation{\KMI}
\author{H.~Iinuma} \affiliation{\IbarakiUniv}
\author{Y.~Ikedo} \affiliation{\KEK}
\author{Y.~Imai} \affiliation{\OkayamaUniv}
\author{K.~Inami} \affiliation{\NagoyaUniv}\affiliation{\KMI}
\author{K.~Ishida} \affiliation{\KEK}
\author{S.~Kamal} \affiliation{\UBC}
\author{S.~Kamioka} \email{kamioka@post.kek.jp} \affiliation{\KEK}
\author{N.~Kawamura} \affiliation{\KEK}
\author{M.~Kimura} \affiliation{\KEK}
\author{A.~Koda} \affiliation{\KEK}
\author{S.~Koji} \affiliation{\NagoyaUniv}
\author{K.~Kojima} \thanks{Present address: \KEK} \affiliation{\KMI}
\author{A.~Kondo} \affiliation{\NagoyaUniv}
\author{Y.~Kondo} \affiliation{\JAEA}
\author{M.~Kuzuba} \affiliation{\IbarakiUniv}
\author{R.~Matsushita} \affiliation{\UnivofTokyo}
\author{T.~Mibe} \affiliation{\KEK}
\author{Y.~Miyamoto} \affiliation{\OkayamaUniv}
\author{J.~G.~Nakamura} \affiliation{\KEK}
\author{Y.~Nakazawa} \thanks{Present address: \KEK} \affiliation{\IbarakiUniv}
\author{S.~Ogawa} \thanks{Present address: \KEK} \affiliation{\RCAPP}
\author{Y.~Okazaki} \affiliation{\KEK}
\author{A.~Olin} \affiliation{\victoria} \affiliation{\TRIUMF}
\author{M.~Otani} \affiliation{\KEK}
\author{S.~Oyama} \affiliation{\UnivofTokyo}
\author{N.~Saito} \affiliation{\KEK}
\author{H.~Sato} \affiliation{\IbarakiUniv}
\author{T.~Sato} \affiliation{\UnivofTokyo}
\author{Y.~Sato} \affiliation{\NiigataUniv}
\author{K.~Shimomura} \affiliation{\KEK}
\author{Z.~Shioya} \affiliation{\KyushuUniv}
\author{P.~Strasser} \affiliation{\KEK}
\author{S.~Sugiyama} \affiliation{\NagoyaUniv}
\author{K.~Sumi} \email{ksumi@hepl.phys.nagoya-u.ac.jp} \affiliation{\NagoyaUniv}
\author{K.~Suzuki} \affiliation{\KMI}
\author{Y.~Takeuchi} \thanks{Present address: Tsung-Dao Lee Institute \& School of Physics and Astronomy, Shanghai Jiao Tong University, Shanghai 201210, China} \affiliation{\KyushuUniv} 
\author{M.~Tanida} \affiliation{\KyushuUniv}
\author{J.~Tojo} \affiliation{\KyushuUniv}\affiliation{\RCAPP}
\author{K.~Ueda} \affiliation{\NagoyaUniv}
\author{S.~Uetake} \affiliation{\OkayamaUniv}
\author{X.~H.~Xie} \affiliation{\PekingUniv}\affiliation{\SKLNPT}
\author{M.~Yamada} \affiliation{\KyushuUniv}
\author{S.~Yamamoto} \affiliation{\OkayamaUniv}
\author{T.~Yamazaki} \affiliation{\KEK}
\author{K.~Yamura} \affiliation{\NiigataUniv}
\author{M.~Yoshida} \affiliation{\KEK}
\author{T.~Yoshioka} \affiliation{\RCAPP}\affiliation{\KyushuUniv}
\author{M.~Yotsuzuka} \affiliation{\NagoyaUniv}
\date{\today}

\begin{abstract}
Acceleration of positive muons from thermal energy to 100~keV has been demonstrated.
Thermal muons were generated by resonant multiphoton ionization of muonium atoms emitted
from a sheet of laser-ablated aerogel.
The thermal muons were first electrostatically accelerated to 5.7~keV, followed by further acceleration to 100 keV using a radio frequency quadrupole with an intensity of $2\times10^{-3}$~$\mu^+$/pulse.
The transverse normalized rms emittance of the accelerated muons in the horizontal and vertical planes were $0.85 \pm 0.25 
 ~\rm{(stat.)}~^{+0.22}_{-0.13} ~\rm{(syst.)}$ ~$\pi~$mm$\cdot$mrad and $0.32\pm 0.03~\rm{(stat.)} ^{+0.05}_{-0.02} ~\rm{(syst.)} $~$\pi~$mm$\cdot$mrad, respectively. The measured emittance values demonstrated phase-space reduction by a factor of $2.0\times 10^2$ (horizontal) and $4.1\times 10^2$(vertical)
allowing good acceleration efficiency.
These results pave the way to realize the first-ever muon accelerator for a variety of applications
in particle physics, material science, and other fields.
\end{abstract}

\maketitle

\begin{figure*}[t]
    \centering\includegraphics[width=.8\linewidth]{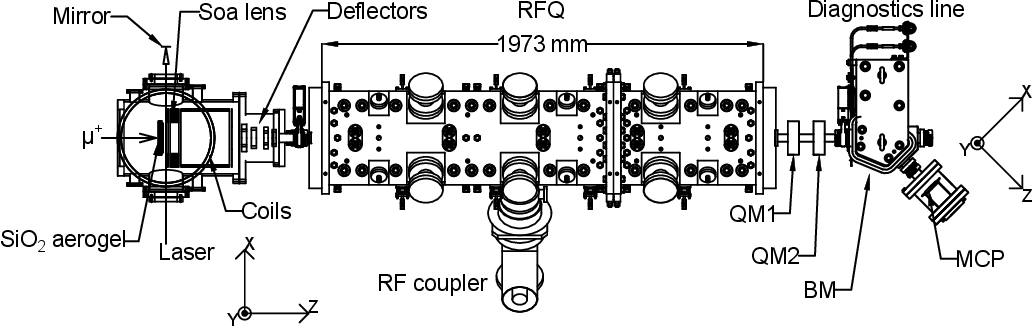}
    \caption{
    Top view of the experimental setup.
    The surface muon beam from the MUSE S line is stopped inside a SiO$_2$ aerogel target. The muonium atoms emitted from the target are ionized by an ionization laser to produce ultraslow muons. The laser travels horizontally and at a 2~mm distance from the target and is reflected by a mirror for the Doppler-free excitation. The ultraslow muons are transported by the Soa lens at 5.7~keV and accelerated to 100~keV by an RFQ. Muons passing through a diagnostic line are detected by an MCP. 
    }
    \label{fig:setup}
\end{figure*}

An accelerated high-intensity positive muon beam enables a variety of next-generation muon experiments.
One such application is a new experiment to measure the muon anomalous magnetic moment and search for the muon electric dipole moment~\cite{10.1093/ptep/ptz030}. 
A transmission muon microscope has also been proposed for high-resolution imaging of thick materials, which is not achievable with conventional transmission electron microscopes~\cite{TmuM}. A $\mu^+e^-$ or $\mu^{+}\mu^{+}$ collider is another promising application for the energy frontier in the future~\cite{10.1093/ptep/ptac059}. 

A high-intensity muon beam with more than 10$^8$~muons/s can be produced from the decay of pions generated in a pion production target irradiated by a megawatt-class proton beam. However, as a tertiary beam, the muon beam initially has a large six-dimensional phase-space volume. Cooling the beam to increase its phase-space density is necessary to realize an efficient acceleration of muons. On the other hand, applying traditional cooling techniques, such as synchrotron radiation cooling, laser cooling~\cite{HANSCH197568}, stochastic cooling~\cite{RevModPhys.57.689}, and electron cooling~\cite{G_I_Budker_1978}, to muons is not feasible due to the short lifetime of muons. 

The cooling techniques dedicated to muons have been intensively developed to realize a muon beam with a small phase-space volume while keeping its intensity high.
The ionization cooling of both positive and negative muons was proposed for a muon collider~\cite{Neuffer:1983jr}. The transverse and longitudinal momentum of the muon beam is lost inside an absorber through ionization, and the longitudinal momentum is restored using radio frequency cavities. Cooling positive muons inside cryogenic helium gas combined with a complex electromagnetic field was also proposed~\cite{PhysRevLett.97.194801}. This technique aims to improve the phase-space density by a factor of 10$^{10}$ with the efficiency of 10$^{-3}$.
These cooling techniques have been demonstrated recently~\cite{Bogomilov2020,Bogomilov2024,PhysRevLett.125.164802}. However, the subsequent acceleration of muons after the cooling has not been reported so far.

Another technique of muon cooling and acceleration has been under development at the muon science facility (MUSE)~\cite{Miyake_2010} of the Japan Proton Accelerator Research Complex (J-PARC). This technique cools positive muons down to the thermal energy of 25~meV and accelerates them using radio frequency (rf) cavities. The thermal muon produced from this technique is called an ultraslow muon. The ultraslow muons are generated through resonant multiphoton ionization of thermal muonium atoms ($\mu^+e^-$) emitted from a muonium production target irradiated by a conventional muon beam~\cite{PhysRevLett.74.4811, BAKULE2008335}. Among the various methods, this technique can produce a muon beam with one of the lowest energies. The ultraslow muons are extracted and transported to linear accelerators by an electrostatic field. 
The cooling efficiencies more than 10$^{-3}$ are achievable with high-energy vacuum ultraviolet and ultraviolet lasers~\cite{10.1093/ptep/ptz030}. 
The transverse phase-space distribution is determined by the spatial profile of the incident muon beam and the velocity distribution of these thermal muons. Normalized transverse root-mean-square (rms) emittance, representing the extent of the phase-space distribution in the transverse plane, as low as 0.3~$\pi$~mm$\cdot$mrad is possible~\cite{10.1093/ptep/ptz030}. 
The pulse duration of the beam is on the order of 1~ns, as determined by the laser pulse duration. The energy spread is $\mathcal{O}(10)$~eV, arising from the spatial spread of the beam and electric field during the extraction. Because of the simple extraction of the cooled muons, this technique is a promising approach to accelerate the cooled muons. 

In this Letter, the first rf acceleration of ultraslow muons to 100~keV is reported, achieved by integrating a muonium target, a low ionization efficiency but precise laser, and a linear accelerator. The transverse emittance of the accelerated muon beam is also evaluated using the quadrupole scan method~\cite{qscan}. 

The experiment reported here was conducted in the S2 area, one of the experimental areas of the surface muon beamline (S line) at MUSE, during March and April of 2024. 
Figure~\ref{fig:setup} shows the experimental setup, consisting of a silica aerogel target for muonium production, an ionization laser, an electrostatic lens (Soa lens), electrostatic deflectors, a radio frequency quadrupole (RFQ), and a beam diagnostic line. 

The MUSE facility provides a pulsed surface muon beam produced by $\pi^{+}$ decay near the surface of the production target. A 3~GeV proton beam was delivered from the rapid-cycling synchrotron operated at a repetition rate of 25~Hz. In this experiment, the proton beam power at the Materials and Life Science Experimental Facility was 940~kW, and the $\mu^+$ intensity was $9.5\times 10^{4}$ muons per pulse ($\mu^+$/pulse). 
The profile of the surface muon beam was measured by a muon beam monitor with a gated image intensifier~\cite{ITO20141}.
The horizontal and vertical rms beam sizes were 24 and 13~mm, respectively. The mean and the rms value of the momentum were 28 and 0.46~MeV/$c$, respectively, based on the simulation of the beam transport using the \texttt{G4BEAMLINE} software~\cite{4440461}. The normalized transverse rms emittances ($\varepsilon_\mathrm{norm,rms}$) in the horizontal and vertical planes were $(1.7 \pm 0.6 )\times 10^2~\pi$ and $ (1.3\pm0.1)\times10^2~\pi$~mm$\cdot$mrad, respectively, based on the simulation. 

\begin{figure*}[t]
    \centering
    \includegraphics[width=0.8\linewidth]{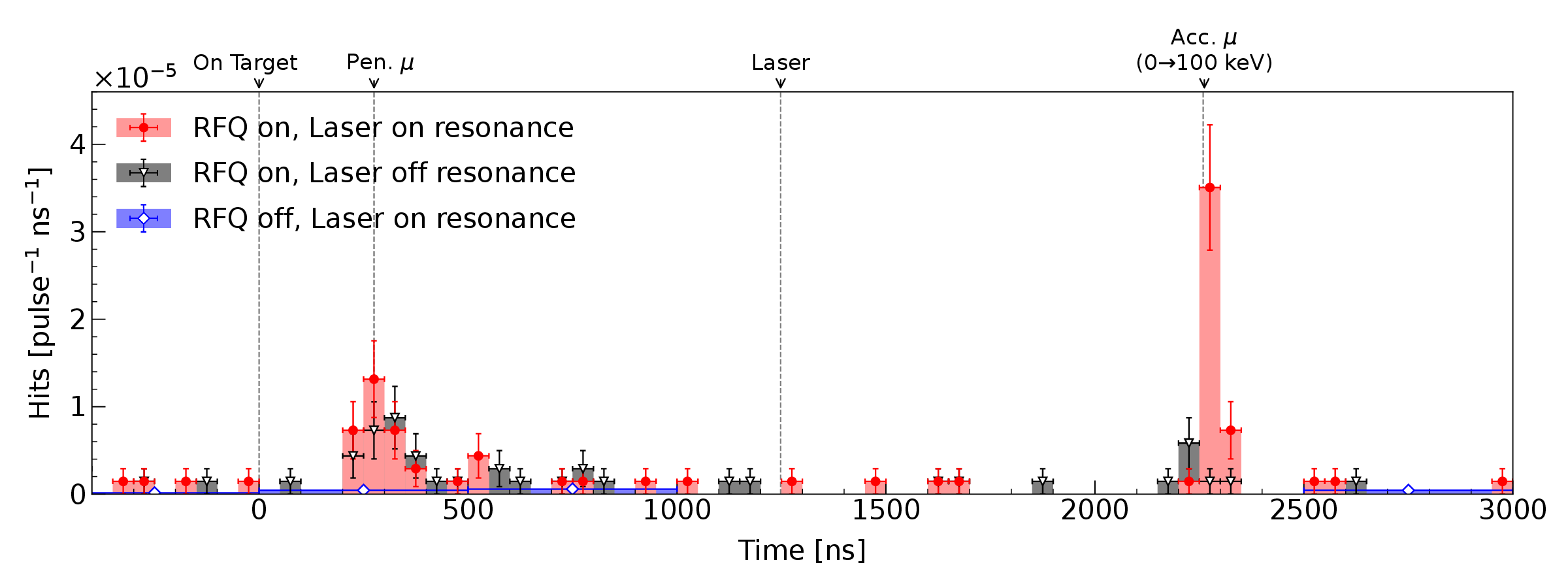}
    \caption{
    TOF distributions of the MCP signal pulse under three conditions: without rf power, with rf power on but the laser off resonance, and with rf power and the laser on resonance. The peak at 2280~ns corresponds to the accelerated muon, while the peak at 275~ns corresponds to the surface muon penetrating the aerogel target. The vertical dashed lines, from left to right, represent the time for the surface muon arriving at the aerogel target, the expected time for the muon penetrating the target to arrive at the MCP, the laser irradiation time, and the expected time for the accelerated muon to arrive at the MCP.
    }
    \label{fig:tof}
\end{figure*}

The surface muons were decelerated by a 420~$\mu$m thick aluminum foil and then incident on a silica aerogel target. The target had a diameter of 78~mm, a thickness of 8~mm, and a density of 23.6~mg/cm$^3$. Both sides of the target were ablated by a short pulse laser, creating 250~$\upmu$m diameter holes at 300~$\upmu$m intervals~\cite{10.1093/ptep/ptu116,10.1093/ptep/ptaa145}. A mesh electrode was placed to the downstream surface of the target for the extraction of the ionized muons.
The stopping efficiency of the surface muon inside the target was 38\% evaluated by a simulation based on 
\texttt{Geant4}~\cite{AGOSTINELLI2003250}. The conversion efficiency from muon to muonium was 52\%~\cite{10.1093/ptep/ptt080}. The muonium emission efficiency from the target was 13\% from a simulation based on a three-dimensional random walk model of muonium diffusion inside the target~\cite{ZHANG2022167443}. Of the emitted muonium atoms, 19\% were lost at the mesh electrode.

A narrow band pulsed light source operated at a wavelength of 244~nm, originally developed for the muonium $1S$-$2S$ spectroscopy experiment~\cite{Zhang:2021cba}, was introduced to first excite the muonium atoms to the 2$S$ state and then ionize them. The light source was based on an injection-seeded titanium-sapphire laser operated at a wavelength of 976~nm. The titanium-sapphire crystals were pumped by an all-solid-state light source operated at a wavelength of 532~nm. The energy of the 244~nm pulse was 0.5--1.3~mJ, depending on the alignment and damage of the optics. The 1/$e^2$ beam diameter was 1~mm, determined by fitting the laser profile to a Gaussian distribution. The pulse duration was 57~ns. The pulse traveled horizontally and parallel to the aerogel surface at a 2~mm distance and was reflected by a mirror for Doppler-free $1S$-$2S$ transition. 
The spatial distribution of the ultraslow muons was highly anisotropic as the beam size along the vertical axis was determined by the laser spot size, while the beam size along the horizontal axis was determined by that of the surface muon beam. The ionization efficiency was defined as the ratio of the ionized muonium atoms to the total emitted muonium atoms, including the muon decay loss after the emission. The ionization efficiency was calculated by solving the optical-Bloch equations. Based on simulations, the efficiency was $7\times10^{-6}$ -- $6\times10^{-5}$, depending on the pulse energy.

The ultraslow muons were accelerated to 5.7~keV and focused at the entrance of the RFQ by an electrostatic lens system, a Soa lens~\cite{soalens}. The extraction energy and focal length were determined to match the acceptance of the following acceleration using the \texttt{musrSim}~\cite{SEDLAK201261}, which is based on \texttt{Geant4}. 
From the simulation, $\varepsilon_\mathrm{norm,rms}$ of the 5.7~keV muon beam were 0.2~$\pi$~mm$\cdot$mrad in the horizontal plane and $2\times10^{-3}$~$\pi$~mm$\cdot$mrad in the vertical plane, respectively. The energy spread was 0.6~eV in 1$\sigma$ from the simulation. 
Three pairs of Helmholtz coils were introduced to compensate for the ambient magnetic field around the muonium target in three directions. Four sets of electrostatic deflectors were used to correct the muon trajectory deflected by the ambient magnetic field. 
Based on a simulation, 20\% of the muons were lost by the muon decay during the transport.

The intensity of the 5.7~keV muon before the rf acceleration was $( 4.0 
\pm 0.4  )\times10^{-3}$~$\mu^+$/pulse. 
For this measurement, a slow muon beamline~\cite{BAKULE2008335} was installed after the Soa lens instead of the deflector and downstream components. The laser frequency was scanned and tuned to excite the $1S-2S$ transition of muonium in the $F=1$ state. The laser irradiation timing was also scanned to maximize the intensity of the muon and fixed to 1.25~$\mu$s after the arrival time of the surface muons.

A prototype RFQ of the J-PARC linac~\cite{hasegawa:linac06-thp072} was used for the acceleration of the ultraslow muons. The RFQ was designed to accelerate negative hydrogen ions to 0.8~MeV. This RFQ also accelerated negative muonium ions to 89~keV~\cite{PhysRevAccelBeams.21.050101,PhysRevAccelBeams.24.033403}. The acceptance of the RFQ in the transverse plane was 3~$\pi~$mm$\cdot$mrad in total normalized emittance. The energy acceptance was 450~eV in 1$\sigma$. 
The forward rf frequency, peak power, and pulse width were set to 324.03~MHz, 2.6~kW, and 40~$\mu$s, respectively.
For the beam dynamics simulation of the RFQ, \texttt{PARMTEQM}~\cite{RFQcode} was used. 
Based on a simulation, 19\% of the muons were lost during acceleration. The transmittance of the RFQ excluding the muon decay was 99\%.

The transverse beam properties of accelerated muons were measured using a beam diagnostics line after the RFQ~\cite{NAKAZAWA2019164}. The beam was transferred using two quadrupole magnets (QM1 and QM2) with field gradients of $2.25\times I_i~(\mathrm{T/m})$, where $I_{i=1,2}~(\mathrm{A})$ is the current supplied to each QM. 
The charge and momentum of the particle were selected using a horizontal bending magnet (BM). 
The beam transportation was simulated using \texttt{General Particle Tracer}~\cite{GPTcode}. The beam loss due to the muon decay was 3\%. No diagnostic for the longitudinal beam properties were implemented.

A single-anode microchannel plate (MCP) located at the downstream end of the beam diagnostics line was used to measure the arrival time of the particles. Its effective area corresponded to a circle 42~mm in diameter, with an aperture ratio of 60\%. Its output signal was digitized with a 250~MS/s waveform digitizer. 
For the profile measurement, an MCP-based beam profile monitor (BPM) was used~\cite{KIM201822}. The effective area of the BPM was 40~mm in diameter. The exposure time was 500~ns for each muon pulse. Signals from muons were identified based on the MCP response, allowing rejection of decay positron events~\cite{OTANI2019162475}.

Figure~\ref{fig:tof} shows the time distributions of the MCP pulses where the time origin is the arrival of the surface muon at the muonium target.
A clear peak was observed at 2280~ns when both the rf power was turned on, and the laser frequency was tuned to the resonance of the $1S$-$2S$ transition, corresponding to the accelerated muons. 
The time of flight (TOF), defined as the time difference between the laser irradiation and the muon arrival at the MCP, was $1032\pm4\pm18$~ns, where the first uncertainty was statistical and the second was systematic, mainly coming from the uncertainty of the laser irradiation time. The TOF agrees with the simulated value of $1009\pm5$~ns, where the uncertainty came from the muon transportation inside the Soa lens. The muon intensity was $ 2 \times10^{-3}$~$\mu^+$/pulse, with 10\% -- 40\% smaller pulse energy of the laser compared to the measurement of the 5.7~keV muon intensity. The acceleration efficiency, defined as the accelerated muon intensity to the 5.7~keV muon intensity excluding the muon decay loss, was 50\%. However, considering the difference in the laser energy between these two measurements, the actual efficiency exceeded this value. 

\begin{figure}
    \centering
    \includegraphics[width=.85\linewidth]{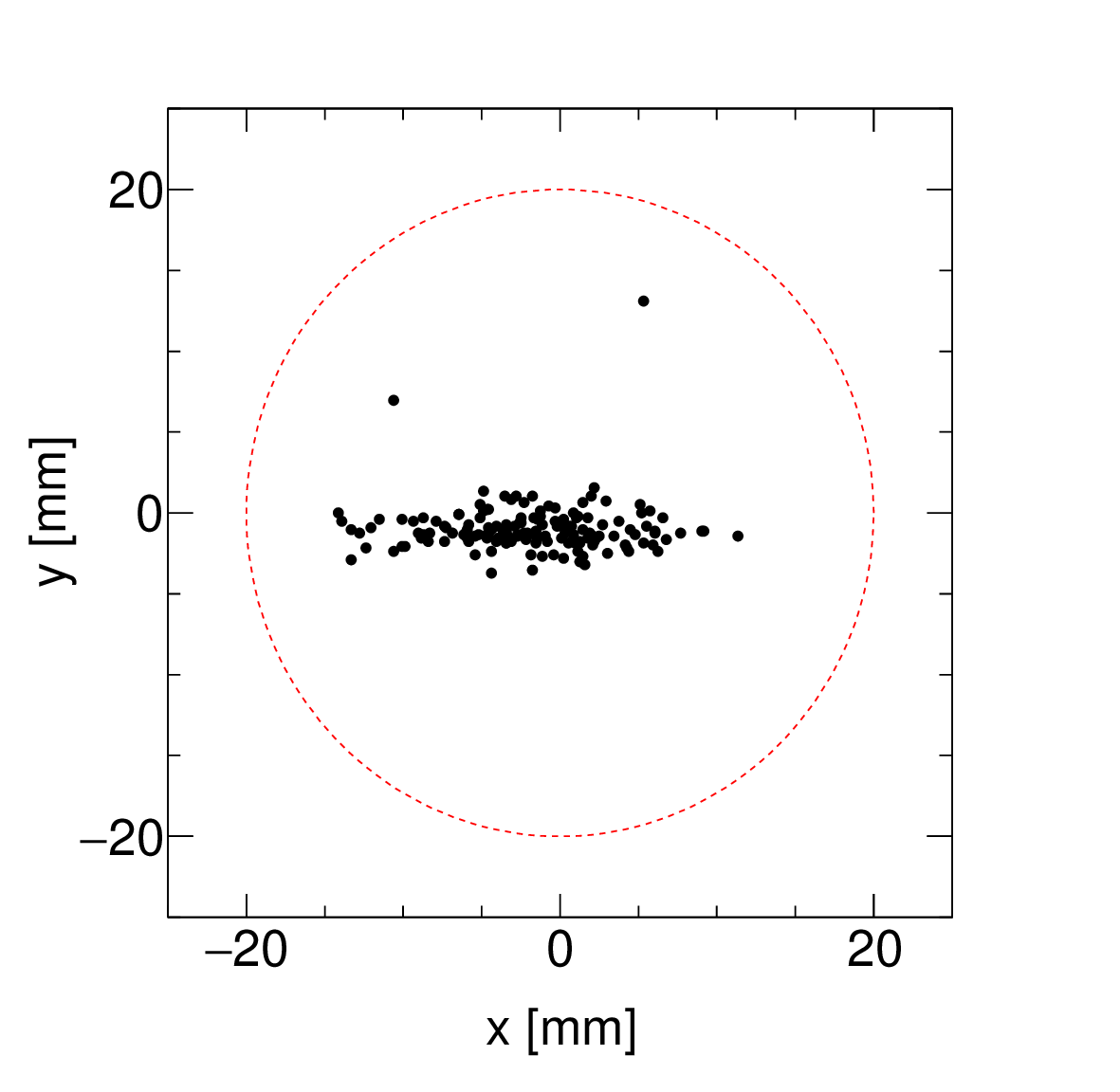}
    \caption{Beam profile measured using a BPM at the end of the diagnostics line. The circle drawn with a dashed line is the BPM’s sensitive area, $\phi$40~mm. The currents of the quadrupole magnets were $I_1 = 0.90$~A and $I_2 = -0.75$~A. 
    }
    \label{fig:profile}
\end{figure}

The measurement of $\varepsilon_\mathrm{norm,rms}$ for the accelerated muon beam 
was conducted using the quadrupole scan method, which determines the transverse phase-space distribution of the beam with a quadrupole magnet and a beam profiler~\cite{qscan}. The rms beam sizes after the BM were measured for different quadrupole strengths of QM2. Figure~\ref{fig:profile} shows an example of the muon beam profile when $I_1=0.9$ and $I_2=-0.75$~A, measured during the quadrupole scan for $\varepsilon_\mathrm{norm,rms}$ in the vertical plane. The rms beam sizes along the horizontal ($x$) and vertical ($y$) axes were calculated from the signal distribution. The rate of background events was evaluated under the laser off-resonance condition. 
Although the origin of the background events remains unidentified, the signal-to-noise ratio was evaluated to be approximately 50 and included in the rms calculation. 
For the profile shown in Fig.~\ref{fig:profile}, the rms beam size along the $y$ axis was $0.87 \pm 0.05 ^{+0.06}_{-0.10}$~mm, where the first uncertainty was statistical and second systematic. The systematic uncertainty, originating from the uncertainty in the number of background events, was evaluated using a Monte Carlo simulation. 

\begin{figure}
    \centering
    \begin{minipage}{.98\linewidth}
    \centering
    \includegraphics[width=.9\linewidth]{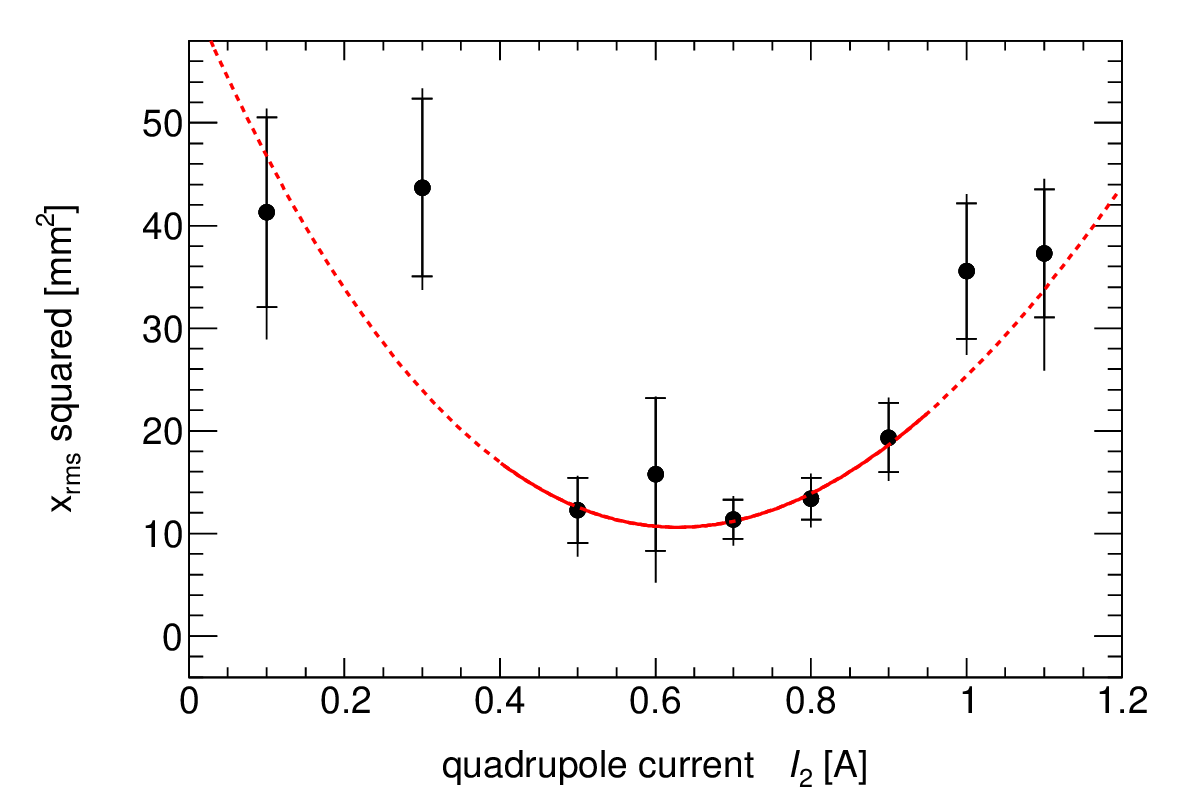}
    \end{minipage}
    \\
    \begin{minipage}{.98\linewidth}
    \centering
    \includegraphics[width=.9\linewidth]{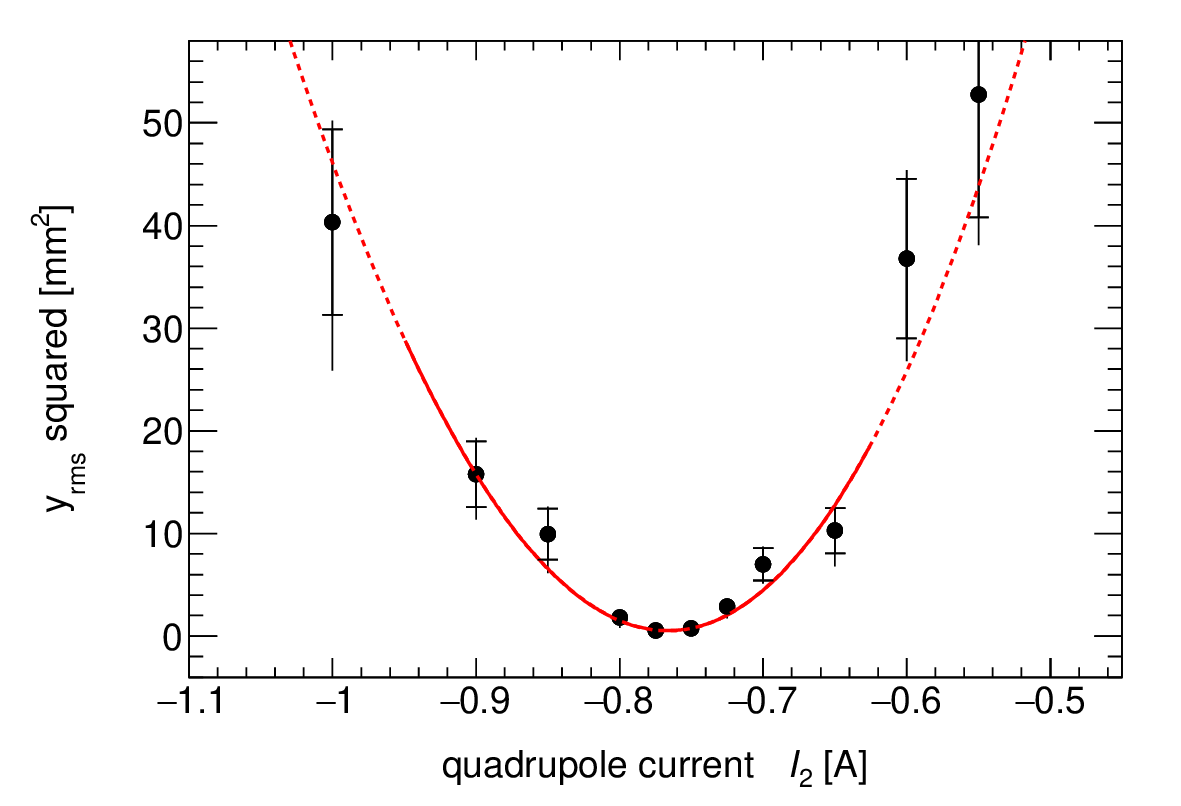}
    \end{minipage}
  \caption{Top: the result of the quadrupole scan measurement along the horizontal direction. Black circles show the square of the rms size of the muon beam. The vertical error bars show total uncertainty, with the horizontal lines indicating the statistical uncertainty. The red solid line shows the fitting result using a transfer function for the region where $x_\mathrm{rms}<5$~mm, while the red dashed line represents the extrapolation of the fitting result to the entire range. Bottom: the result in the vertical plane.}
    \label{fig:q-scan}
\end{figure}

Figure~\ref{fig:q-scan} shows the results of the quadruple scan. 
The normalized rms emittance, $\varepsilon_\mathrm{norm,rms}$, was estimated by fitting them using the transfer matrix, which included QM1, QM2, BM, and the drift space. Data points corresponding to beam sizes larger than 5 mm were excluded from the fit, as larger beam sizes could introduce bias in the rms calculation due to detector size limitations. The momentum spread of the muon beam, evaluated to be $dp/p = (1.10^{+0.02}_{-0.11})\times10^{-2}$ from \texttt{PARMTEQM}, was included in the fitting function to account for the dispersion. The uncertainty of the momentum spread originated from the rf power evaluation.

Systematic uncertainties on $\varepsilon_\mathrm{norm,rms}$ are shown in Table~\ref{table:Tab1}. The uncertainty from the momentum spread was assessed by varying its value in the fitting function. The uncertainty from beam loss in the diagnostic line was evaluated through a simulation.

Finally, the measured $\varepsilon_\mathrm{norm,rms}$ of the accelerated muon beam in the horizontal and vertical planes were $(0.85 \pm 0.25 ^{+0.22}_{-0.13} )$~$\pi$ and $(0.32\pm 0.03  ^{+0.05}_{-0.02})$~$\pi~$mm$\cdot$mrad, respectively. 
The simulated $\varepsilon_\mathrm{norm,rms}$ in the horizontal and vertical planes were $(0.27 \pm 0.04)~\pi$ and $(0.154\pm0.008)$~$\pi$~mm$\cdot$mrad, respectively. 
The discrepancy between measured and simulated emittance suggests the presence of an unaccounted mismatch factor in the beam transport optics within the simulation.


\begin{table}
    \centering
    \caption{Systematic uncertainties on $\varepsilon_\mathrm{norm,rms}$.}
        \begin{tabular}{lWc{30mm}Wc{20mm}}
        \hline \hline
         \multirow{2}{*}{Source} &  \multicolumn{2}{c}{Contribution ($\pi$~mm$\cdot$mrad)} \\
         & Horizontal & Vertical \\
        \midrule
        Rms beam size    &  $\pm0.13$  &  $\pm0.02$  \\
        Momentum spread   &  $^{+0.12}_{-0.02}$  & --       \\
        Beam loss   &         $+0.14$  &   $+0.05$      \\
        \midrule
        Total   &     $^{+0.22}_{-0.13} $  &  $^{+0.05}_{-0.02}$  \\
        \hline \hline
        \end{tabular}
    \label{table:Tab1}
\end{table}

In summary, muons were cooled and accelerated by an rf accelerator for the first time. The ultraslow muons were generated by the resonant multiphoton ionization of muonium atoms emitted from a laser-ablated aerogel target using an all-solid-state laser at 244~nm, and accelerated to 100~keV. The intensity of the muon beam was $2\times10^{-3}$~$\mu^+$/pulse. Compared to the surface muon beam, the normalized transverse rms emittance was reduced by a factor of $2.0\times 10^2$ (horizontal) and $4.1\times 10^2$ (vertical), demonstrating the achievement of a low emittance muon beam through the acceleration of a cooled muon by an rf cavity with a good acceleration efficiency. 

Toward the application of this muon beam to various experiments, increasing its intensity is essential. The low muon rate in this experiment was due to the choice of a laser optimized for spectroscopy rather than ionization. By developing high-energy light sources operated at 122 and 355~nm dedicated to the ionization of muonium, the ionization efficiency is improved to around 10\%~\cite{10.1093/ptep/ptz030}. The light source is currently under development at J-PARC~\cite{Saito:16}. In addition, a new muon beamline to deliver 70 times more intense surface muon beam is under construction~\cite{10.1093/ptep/pty116}. In total, 10$^5$ -- 10$^6$ increase of the muon rate is expected, potentially enabling the applications such as the muon $g-2$/EDM experiment at J-PARC~\cite{10.1093/ptep/ptz030}.

The results presented in this Letter show the phase-space reduction of a muon beam by more than 2 orders of magnitude. This enables broad applications of an accelerated positive muon beam from particle physics to various other scientific fields.

\begin{acknowledgments}
The authors would like to thank the J-PARC muon section
staff for their support in the conduct of the experiment at
J-PARC MUSE. This work was supported in part by JSPS Kakenhi Grants No. JP18H05226, No. JP19H05606, No. JP20H05625, No. JP21K13944, No. JP21J01132, No. JP22KJ1594, No. JP22K21350, No. JP22H00141, No. JP24H00023, No. JP24K03211, MEXT Q-LEAP JPMXS0118069021, and JST-Mirai Program JPMJMI17A1. The experiment was performed at the Materials and Life Science Experimental Facility of the J-PARC under a user program 2011MS06.
\end{acknowledgments}


\bibliography{main}

\end{document}